\documentclass[twocolumn,aps,amssymb,floatfix,superscriptaddress,preprintnumbers]{revtex4}

\usepackage{graphicx}
\usepackage{bm}
\usepackage{amsmath}
\usepackage{amssymb}
\usepackage{amsfonts}
\usepackage{float}
\usepackage{hyperref}
\usepackage{dsfont}  
\usepackage{slashed}  
\usepackage{booktabs}
\usepackage{multirow}
\usepackage{subfigure}
\usepackage[sort&compress]{natbib}
\usepackage{xcolor}
\usepackage{ulem}

\newcommand{\be}{\begin{equation}}  
\newcommand{\ee}{\end{equation}}  
\newcommand{\beq}{\begin{eqnarray}}  
\newcommand{\eeq}{\end{eqnarray}}

\newcommand{\bea}{\begin{eqnarray}}
\newcommand{\eea}{\end{eqnarray}}

\newcommand{\MSb}{{\overline{\rm MS}}}

\begin{document}

\title{New insights on proton structure from lattice QCD: \\ the twist-3 parton distribution function $g_T(x)$}

\author{Shohini Bhattacharya}
\affiliation{Department of Physics,  Temple University,  Philadelphia,  PA 19122 - 1801,  USA}
\author{Krzysztof Cichy}
\affiliation{Faculty of Physics, Adam Mickiewicz University, ul.\ Uniwersytetu Pozna\'nskiego 2, 61-614 Pozna\'{n}, Poland}
\author{Martha Constantinou}
\affiliation{Department of Physics,  Temple University,  Philadelphia,  PA 19122 - 1801,  USA}
\author{Andreas Metz}
\affiliation{Department of Physics,  Temple University,  Philadelphia,  PA 19122 - 1801,  USA}
\author{Aurora Scapellato}
\affiliation{Faculty of Physics, Adam Mickiewicz University, ul.\ Uniwersytetu Pozna\'nskiego 2, 61-614 Pozna\'{n}, Poland}
\author{Fernanda Steffens}
\affiliation{Institut f\"ur Strahlen- und Kernphysik, Rheinische
  Friedrich-Wilhelms-Universit\"at Bonn, Nussallee 14-16, 53115 Bonn}

\begin{abstract}

\vspace*{1cm}
\noindent In this work, we present the first-ever calculation of the isovector flavor combination of the twist-3 parton distribution function $g_T(x)$ for the proton from lattice QCD. We use an ensemble of gauge configurations with two degenerate light, a strange and a charm quark ($N_f=2+1+1$) of maximally twisted mass fermions with a clover improvement. The lattice has a spatial extent of 3~fm, lattice spacing of 0.093~fm, and reproduces a pion mass of $260$ MeV. We use the quasi-distribution approach and employ three values of the proton momentum boost, 0.83 GeV, 1.25 GeV, and 1.67 GeV. We use a source-sink separation of 1.12~fm to suppress excited-states contamination. The lattice data are renormalized non-perturbatively. 
We calculate the matching equation within Large Momentum Effective Theory, which is applied to the lattice data in order to obtain $g_T$. 
The final distribution is presented in the $\MSb$ scheme at a scale of 2 GeV. 
We also calculate the helicity distribution $g_1$ to test the Wandzura-Wilczek approximation for $g_T$. We find that the approximation works well for a broad range of $x$.

\medskip
\noindent This work demonstrates the feasibility of accessing twist-3 parton distribution functions from novel methods within lattice QCD and can provide essential insights into the structure of hadrons.
  
\end{abstract}

\maketitle

\noindent
\textit{Introduction:}
More than 99\% of the mass of the visible world resides in atomic nuclei and, therefore, in nucleons, that is, protons and neutrons.
Nucleons, in turn, are complicated bound states of quarks and gluons (partons), which are the fundamental degrees of freedom of quantum chromodynamics (QCD), the microscopic theory of the strong interaction. 
Trying to understand the rich parton structure of nucleons has been among the most important and active research areas in hadronic physics for several decades. 

The first experimental evidence of a partonic substructure of the proton emerged from measurements of deep-inelastic electron-proton scattering (DIS), $e p \to e X$, in the late 1960s~\cite{Bloom:1969kc, Breidenbach:1969kd}.
These experiments were, in fact, instrumental for the discovery of QCD.
The DIS cross-section can be parameterized in terms of four terms (structure functions) --- two for unpolarized initial-state electron and proton, and two if both are polarized, where the latter are often denoted by $g_1^{\rm s.f.}$ and $g_2^{\rm s.f.}$~\cite{Jaffe:1996zw}.
QCD factorization theorems allow one to separate the structure functions into a perturbatively calculable part and a non-perturbative part that contains information about the parton structure of the proton~\cite{Collins:1989gx}.
The non-perturbative contribution is given by parton distribution functions (PDFs), which are, therefore, fundamental quantities characterizing the parton structure of the proton~\cite{Collins:1981uw}.

PDFs can be classified according to their twist, which describes the order in $1/Q$ at which 
they appear in the factorization of the structure functions, with $Q$ denoting the large energy scale of the process. 
(For DIS, $Q$ is the momentum transfer between the initial and final electrons.)
The leading-power PDFs appearing in the factorization are labeled twist-2 PDFs.
They can be considered probability densities for finding, inside the proton, a parton which carries the fraction $x$ of the proton momentum.
Twist-3 PDFs are very important as well.
They are not necessarily smaller than twist-2 PDFs.
While they do not have a density interpretation, twist-3 PDFs contain information about quark-gluon-quark correlations~\cite{Balitsky:1987bk,Kanazawa:2015ajw}, and as such, characterize the structure of hadrons in a new way. 
They appear in QCD factorization theorems for a variety of hard scattering processes and have interesting connections with transverse momentum dependent parton distributions, thus offering essential insights into the latter, see, e.g., Refs.~\cite{Jaffe:1996zw,Cammarota:2020qcw,Accardi:2009au}.

In this work, we present the first-ever calculation of the twist-3 PDF $g_T(x)$ using lattice QCD. This PDF enters the aforementioned structure functions $g_1^{\rm s.f.}$ and $g_2^{\rm s.f.}$. Therefore, our calculation is a crucial step forward to fully understand the DIS process from first principles in QCD. It also complements efforts to extract information about $g_T$ from experiment~\cite{Flay:2016wie, Armstrong:2018xgk}. Generally, measurements related to twist-3 PDFs are part of the ongoing 12 GeV program at Jefferson Lab and will be important for the planned Electron-Ion Collider~\cite{Boer:2011fh,Accardi:2012qut}. However, measuring twist-3 PDFs is difficult due to their (suppressed) ${\mathcal{O}(1/Q)}$ kinematical behavior.

Specifically, we discuss the calculation of the isovector flavor combination $g_T^{u-d}(x)$. (For ease of notation, we omit the superscript $u-d$ in the remainder of this paper.) We make use of the so-called quasi-PDF approach suggested by X.~Ji~\cite{Ji:2013dva,Ji:2014gla}. While standard (light-cone) PDFs are defined through light-cone correlation functions, quasi-PDFs and related quantities~\cite{Braun:2007wv,Radyushkin:2017cyf,Ma:2017pxb} are given by spatial correlation functions accessible in lattice QCD, which gave rise to an intensive surge of studies, see, e.g., Refs.~\cite{Lin:2014zya,Alexandrou:2015rja,Chen:2016utp,Alexandrou:2016jqi,Chambers:2017dov,Alexandrou:2017huk,Orginos:2017kos,Ishikawa:2017faj,Ji:2017oey,Radyushkin:2018cvn,Alexandrou:2018pbm,Chen:2018fwa,Alexandrou:2018eet,Liu:2018uuj,Karpie:2018zaz,Zhang:2018diq,Li:2018tpe,Sufian:2019bol,Karpie:2019eiq,Alexandrou:2019lfo,Izubuchi:2019lyk,Cichy:2019ebf,Joo:2019jct,Radyushkin:2019owq,Joo:2019bzr,Alexandrou:2019dax,Chai:2020nxw,Ji:2020baz,Braun:2020ymy,Bhat:2020ktg,Alexandrou:2020uyt,Bringewatt:2020ixn} and the recent reviews in Refs.~\cite{Cichy:2018mum,Ji:2020ect,Constantinou:2020pek}. Because quasi-PDFs and light-cone PDFs have the same infrared (non-perturbative) physics~\cite{Xiong:2013bka,Ma:2014jla,Briceno:2017cpo,Ma:2017pxb}, they can be related using perturbative QCD, in a procedure called matching, see Refs.~\cite{Stewart:2017tvs,Izubuchi:2018srq,Alexandrou:2019lfo,Zhang:2018diq,Wang:2019tgg,Balitsky:2019krf} for related recent work. The matching equations are known only for twist-2 operators, and within this work we address the one-loop matching kernel for $g_T$.

Our calculation also allows us to address the validity of the Wandzura-Wilczek (WW) approximation for  $g_T(x)$~\cite{Wandzura:1977qf}. The Mellin moments ($x$-moments) of $g_T(x)$ receive contributions from twist-2 operators and twist-3 operators (whose moments we denote by $d_n$). Therefore, $g_T(x)$ can be written as $g_T^{\rm WW}(x) + g^{\rm twist-3}_T(x)$, with $g^{\rm twist-3}_T (x)$ the contribution from twist-3 operators~\cite{Accardi:2009au}. In the WW approximation, one sets $d_n =0$, implying that $g_T(x)$ is fully determined by the twist-2 operators which define twist-2 helicity PDF $g_1(x)$. Thus, the study of the WW approximation gives direct information about the importance of twist-3 operators. Here, we present the first check in lattice QCD of how relevant the twist-3 operators are for the $x$-dependence of $g_T(x)$.

\vspace*{0.25cm}
\noindent \textit{Methodology:}
The calculation is based on matrix elements of a non-local operator, with space-like separated fermion fields, which are connected via a straight Wilson line (WL) of length $z$. The operator has a Dirac structure $\gamma^j\,\gamma^5$, and the matrix element is defined in position ($z$) space as
\begin{equation}
\label{eq:ME}
{\cal M}_{g_T}(P,z)\,=\,\langle P\vert \, \overline{\psi}(0,z)\,\gamma^j\,\gamma^5\, W(z)\,\psi(0,0)\,\vert P\rangle\,.
\end{equation}
The proton is boosted in a spatial direction, and the quasi-distributions approach requires that it is in the same direction as the WL, i.e.\ $P=(i E,0,0,P_3)$. To obtain the twist-3 distribution, $\gamma^j$ must be $\gamma^x$ or $\gamma^y$, each requiring a parity projector $(1+\gamma^0) i \gamma^5 \gamma^j/4$. In this work, we average over the two operators to increase the statistical accuracy.

For the proper evaluation of $g_T$, one must extract the ground-state contribution from ${\cal M}_{g_T}$. This is achieved by a large time separation between the initial (source) and final (sink) state of the proton, $T_{\rm sink}$, as well as by a current insertion that is away from the source and the sink. Once these conditions are satisfied, we identify the ground state using a fit to a constant (plateau region). The desired quantity, ${F}_{g_T}$, is extracted based on the continuum decomposition:
\begin{equation}
{F}_{g_T}(P_3,z) = - i \frac{E}{m} Z_{g_T}(z)\, {\cal M}_{g_T}(P_3,z)\,,
\end{equation}
in Euclidean space. The kinematic factor is obtained based on the normalization conventions on the lattice. $ E $ is the proton's energy, $E=\sqrt{m^2 + P_3^2}$, $m$ is its mass, and $Z_{g_T}$ is the renormalization function, and it is also calculated in this work.

The so-called quasi-distribution, $\widetilde{g}_T$, is defined as the Fourier transform of ${F}_{g_T}(P_3,z)$ over $z$. It is, thus, given in the momentum representation, $x$,
\begin{equation}
\label{eq:quasi_pdf}
\widetilde{g}_T(x,\Lambda,P_3) = 2 P_3 \int_{-\infty}^{+\infty}\hspace*{-0.1cm}\frac{dz}{4\pi}\,
e^{-ixP_3z}\,{ F}_{g_T}(P_3,z) ,
\end{equation}
where $\Lambda{\sim} 1/a$ is a UV cut-off. Our definition of $\widetilde{g}_T$ is such that its lowest $x$-moment is independent of $P_3$, see also Ref.~\cite{Bhattacharya:2019cme}.
As the momentum $P_3$ increases, the quasi-distribution $\widetilde{g}_T$ can be matched to the light-cone distribution $g_T$ using a perturbative formula obtained within Large Momentum Effective Theory \cite{Ji:2013dva,Ji:2014gla}.

\vspace*{0.25cm}
\noindent\textit{Computational setup:}
\noindent In this work, we use one $N_f=2+1+1$ ensemble of twisted mass fermions \cite{Frezzotti:2000nk,Frezzotti:2003ni} with clover improvement \cite{Sheikholeslami:1985ij} and Iwasaki gluons~\cite{Alexandrou:2018egz}. 
The lattice spacing is 0.093 fm, its volume is 32$^3\times$64 ($L\approx3$ fm), and the pion mass is around 260 MeV.
  
We focus on the isovector combination, which receives contributions only from the connected diagram. $T_{\rm sink}$ is taken to be above 1 fm ($T_{\rm sink}=1.12$ fm), for which excited-states contamination is assumed to be suppressed for the values of $P_3$ we employ~\cite{Alexandrou:2019lfo}. We apply stout smearing~\cite{Morningstar:2003gk} to the links of the operator, which is known to reduce statistical uncertainties in matrix elements of non-local~\cite{Alexandrou:2019lfo}, and gluonic~\cite{Alexandrou:2016ekb,Alexandrou:2020sml} operators.

In this work, ${\cal M}_{g_T}$ is calculated for three values of momentum boost, $P_3=0.83,\,1.25,\,1.67$ GeV. The statistical uncertainties increase with the momentum, and therefore, the number of measurements must increase by around one order of magnitude with each additional momentum unit to achieve similar statistical errors. We use the momentum smearing method~\cite{Bali:2016lva} on the proton interpolating field, which offers a better overlap of the interpolator and the ground state. The momentum smearing parameter has been tuned following the procedure described in~\cite{Alexandrou:2019lfo}, and leads to a significant reduction of statistical uncertainties. We achieve similar accuracy for each boost with 1552, 11696, and 105216 measurements at $P_3=0.83$, $1.25$, and $1.67$ GeV, respectively. Using the same ensemble, simulation parameters, and statistics, we also obtain the leading-twist helicity PDF, $g_1(x)$. More details on the lattice calculation can be found in the supplementary material. The dependence of the bare $F_{g_T}$ on $P_3$ at each $z/a$ value is shown and discussed in Fig.~S2.

\vspace*{0.25cm}
\noindent \textit{Renormalization:} 
One of the crucial aspects of the calculation is the renormalization of the bare matrix elements. Non-local operators containing a WL require an evolved renormalization procedure, in contrast to the local fermion operators. The presence of the WL is associated with a power divergence with respect to the lattice spacing~\cite{Dotsenko:1979wR,Brandt:1981kf}. Such divergence must be removed along with all logarithmic divergences so that physical meaning can be attributed to lattice data. Considering that this is the first study of twist-3 distributions and that the main focus on the extraction of the matrix elements, we do not take into account any mixing with other twist-3 operators (e.g., with quark-gluon-quark operators). It is expected that the only non-local UV divergence in the quasi-distributions is the power divergence due to the presence of the Wilson line. This has been discussed and confirmed for the case between the quark singlet and gluon PDFs~\cite{Zhang:2018diq,Wang:2017qyg}. Consequently, the quasi counterpart of $g_T$ does not mix with quark-gluon-quark non-local operators. Such a mixing manifests itself in the matching formalism. It is useful to consider that the effects due to mixing are often small (less than $10\%$), and in many cases, within the reported uncertainties. Such cases are, for example, the mixing in the singlet quark and gluon momentum fractions~\cite{Alexandrou:2020sml}, but also in a more complicated mixing pattern involving multi-parton operators~\cite{Constantinou:2017sgv}.
Of course, much more work is needed to pin down the numerical significance of the mixing nature in the present case. However, based on the above and on the overall size of matching effects (see below), there is an indication that the mixing is within the reported precision of the final results.

We employ the renormalization procedure developed for straight-WL non-local operators~\cite{Constantinou:2017sej,Alexandrou:2017huk}, and is also used for twist-2 distributions (see, e.g., Ref.~\cite{Alexandrou:2019lfo}). We calculate non-perturbatively the renormalization functions in the RI$'$ scheme~\cite{Martinelli:1994ty} at each value of $z/a$ separately. We use a set of five $N_f=4$ ensembles~\cite{Alexandrou:2019ali} produced specifically for the renormalization functions of the ensemble used in this work. The renormalization procedure is outlined in the supplementary material. We eliminate possible systematic uncertainties in the renormalization functions by an advanced program, in which we: \textbf{a.} perform a chiral extrapolation on the five ensembles, \textbf{b.} use a wide range of RI$'$ renormalization scales and fit the $\MSb$ estimates to eliminate any dependence on the initial scale, \textbf{c.} remove discretization effects utilizing results in lattice perturbation theory~\cite{Alexandrou:2015sea}. The renormalization factors are complex functions due to the presence of the WL, and thus, the renormalized matrix elements are obtained from the complex multiplication $Z_{g_T}^{\MSb}(z) \cdot{\cal M}_{g_T}(P,z)$. The renormalized matrix elements are given in the modified $\MSb$-scheme (${\rm M}\MSb$)~\cite{Alexandrou:2019lfo} at the scale of 2 GeV.
 
\vspace*{0.25cm}
\noindent\textit{Reconstruction of $x$-dependence:} 
The lattice calculation provides determinations of ${F}_{g_T}(P_3,z)$ for discrete values of $z\leq z_{\rm max}$, with $z_{\rm max}\sim L/2$.
Thus, Eq.~(\ref{eq:quasi_pdf}) needs to be discretized and becomes subject to an ill-defined inverse problem, as discussed in Ref.~\cite{Karpie:2018zaz}.
One of the methods advocated to solve this problem is the Backus-Gilbert method \cite{BackusGilbert}, which maximizes the stability of the solution with respect to the statistical variation of the data.
Thus, it provides a model-independent assumption allowing one to obtain a unique reconstructed quasi-distribution from the available set of matrix element evaluations. We employ the Backus-Gilbert method for the results presented here.
 
\vspace*{0.25cm}
\noindent\textit{Matching to light-cone $g_T(x)$:}
Another novel aspect of this work is the calculation of
\begin{equation}
\label{eq:matching}
g_T(x,\mu)=\int_{-\infty}^\infty 
\frac{d\xi}{|\xi|} \, C\left(\xi,\frac{\mu}{x P_3}\right)\, \widetilde{g}_T\left(\frac{x}{\xi},\mu,P_3\right)\,,  
\end{equation}
which connects $\widetilde{g}_T(x)$ to the light-cone $g_T(x)$. 
$C$ is the matching kernel, which is calculated within one-loop perturbation theory in momentum space. Matching for twist-3 distributions has never been addressed in the literature.
We present here the first matching formula for the twist-3 PDF $g_T$.
We repeat that Eq.~(\ref{eq:matching}) does not take into account mixing with quark-gluon-quark operators, which would change the general structure of the equation.
We explore two schemes for the matching, which use the same bare matrix elements, but the renormalization functions are converted to different schemes. The first one uses $Z^\MSb$, while the second one uses the so-called modified $\MSb$-scheme, $Z^{{\rm M}\MSb}$. The latter preserves the normalization, unlike the $\MSb$ scheme, through an extra renormalization in the ``unphysical'' $|\xi|>1$ region. More details on the perturbative calculation for the matching kernel can be found in a separate publication~\cite{Bhattacharya:2020xlt}.
For the results presented here, we employ the ${\rm M}\MSb$ scheme for the quasi-distributions, for which $C$ takes the form:
\begin{widetext}
\begin{align}
C_{{\rm M}\MSb}\left(\xi,\frac{\mu}{x P_3}\right)
&=\delta(\xi-1)
- \frac{\alpha_s}{2\pi}C_F\left\{\begin{aligned}
    &\left[\frac{\xi^2-2\xi-1}{1-\xi}\ln\frac{\xi-1}{\xi}+\frac{\xi}{1-\xi}+\frac{3}{2\xi}\right]_+  &\xi>1,\\
    &\left[\frac{-\xi^2+ 2\xi+1}{1-\xi}\left(\ln\frac{4(x P_3)^2}{\mu^2} + \ln(\xi(1-\xi))\right) + \frac{\xi^2-\xi-1}{1-\xi}\right]_+  &0<\xi<1,\\
&\left[\frac{-\xi^2+2\xi+1}{1-\xi}\ln\frac{\xi-1}{\xi}-\frac{\xi}{1-\xi}+\frac{3}{2(1-\xi)}\right]_+  &\xi<0.
\end{aligned}\right. \nonumber\\
\end{align}
\end{widetext} 
The numerical effect of the above matching is found to be of similar magnitude as for twist-2 PDFs (see e.g.\ Fig.\ 31 of Ref.~\cite{Alexandrou:2019lfo}).
Note that the light-cone $g_T(x)$ results are always in the $\MSb$-scheme, regardless of the scheme used for $\widetilde{g}_T(x)$. In deriving the matching coefficient we did not consider potential mixing with quark-gluon-quark operators.

\begin{figure}[h!]
 	\hspace*{-0.2cm}    \includegraphics[scale=0.57]{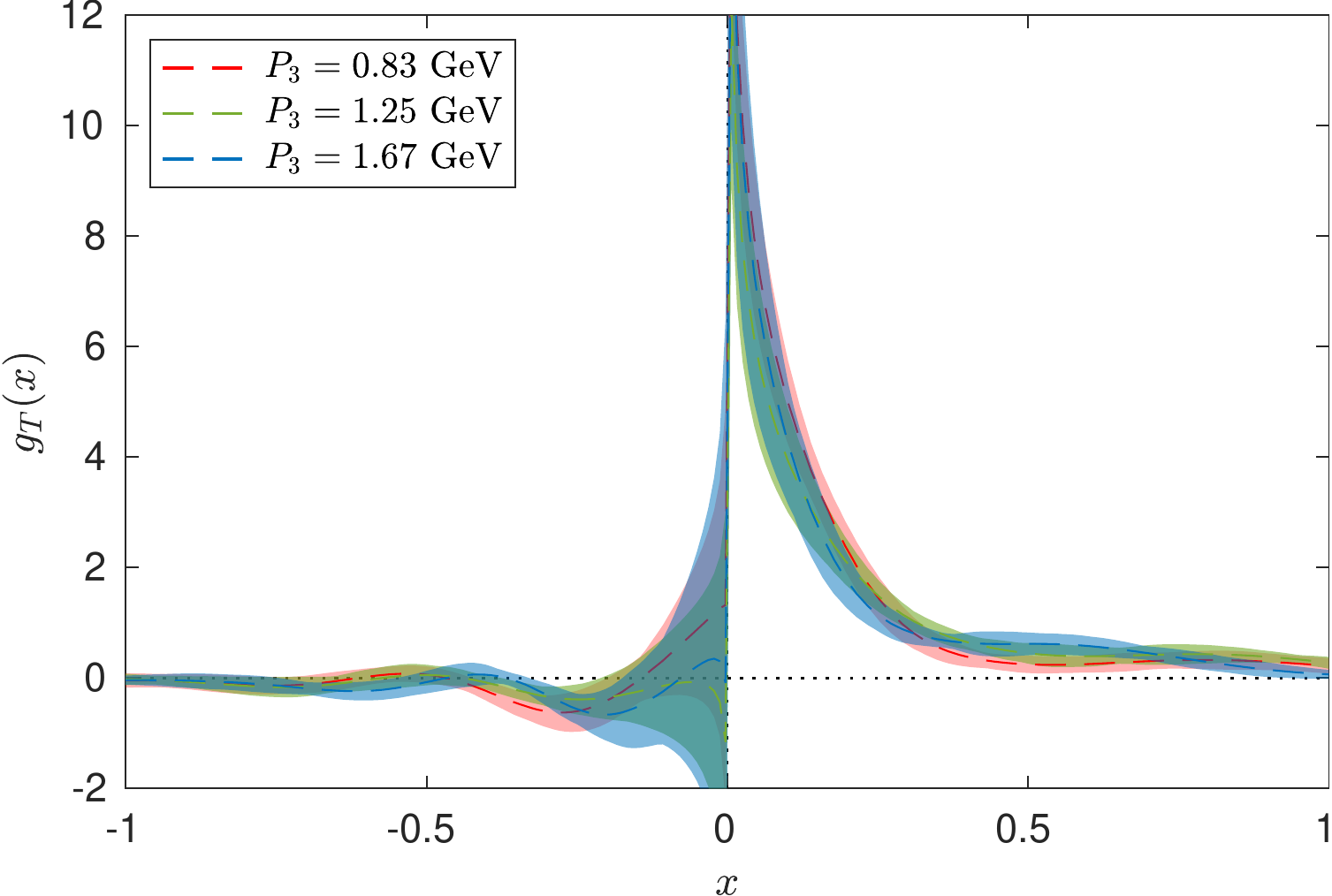}
 	\caption{Results for $g_T(x)$ as a function of $x$ for three nucleon momenta.
 	$P_3=0.83,\,1.25$, and 1.67 GeV are shown with red, green and blue bands, respectively.}
 	\label{fig:gTfinal}
 \end{figure}

\vspace*{0.25cm}
\noindent \textit{Results on $g_T(x)$:}
The various steps described above are combined to provide the final estimates for the twist-3 distribution $g_T(x)$. The renormalized ground-state contributions to the matrix elements, $F_{g_T}$, are transformed to $x$-space using the Backus-Gilbert method and then matched using Eq.~(\ref{eq:matching}). In Fig.~\ref{fig:gTfinal}, we plot the dependence of $g_T(x)$ on the proton momentum $P_3$ for the quark ($x>0$), and antiquark ($x<0$) regions. With red, green and blue bands we show the distributions for $P_3=0.83,\,1.25$, and 1.67 GeV, respectively. The width of each band represents the uncertainties. We consider statistical errors and systematic effects due to the $x$-dependence reconstruction. To account for this, we add an uncertainty, which is computed by varying the maximum value of $z$ entering the reconstruction. The final error is chosen based on the quasi-PDFs' maximum variation using 25 combinations of the $z$ interval. This procedure is applied for both $g_T$ and $g_1$. Statistical and systematic uncertainties are added in quadrature, and these combined errors are used for all the results presented here. 
Note that neglecting the mixing mentioned above with quark-gluon-quark correlators also gives rise to systematic uncertainties in all our numerical results for $g_T$. At present, we do not expect this point to alter any of our general conclusions.
We find that the distribution in the region $x<0.4$ becomes slightly more narrow as the momentum increases, but within the included systematic uncertainties. Besides, we find convergence between the two largest momenta for all regions of $x$. Regarding antiquarks, we observe similar functional forms for these momenta.

\begin{figure}[h!]
  \hspace*{-0.2cm} 
\centering
  \includegraphics[scale=0.58]{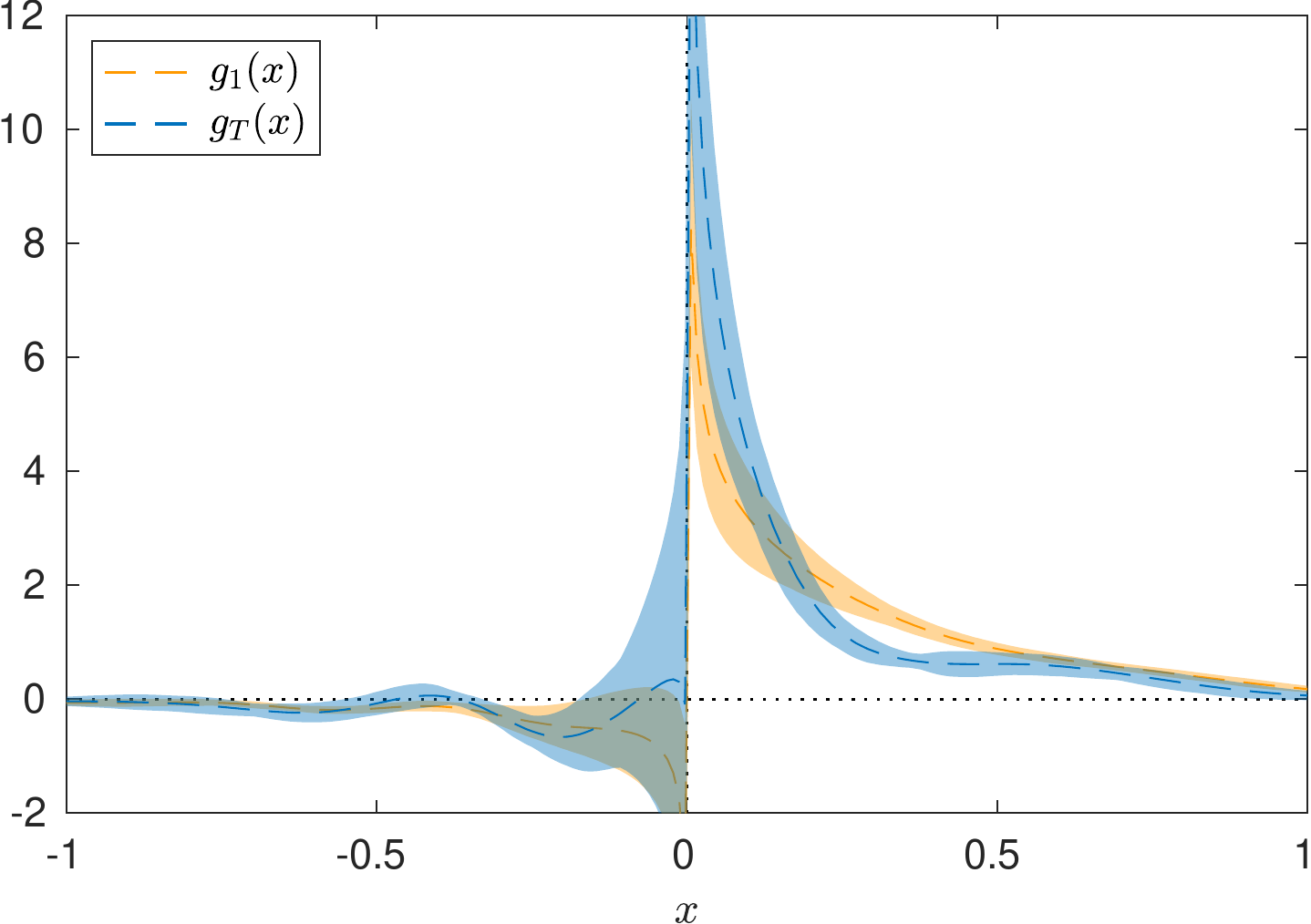}
 	\caption{Comparison of $x$-dependence of $g_T$ (blue band) and $g_1$ (orange band) at $P_3=1.67$ GeV.}
 	\label{fig:gT_g1}
 \end{figure}

Since $g_T$ is a sub-leading contribution, it is interesting to compare it with the leading-twist PDF. To this end, we calculate the twist-2 helicity PDF $g_1(x)$ on the same ensemble and using the same values for $P_3$ and $T_{\rm sink}$. In Fig.~\ref{fig:gT_g1}, we show the results for the highest momentum, $P_3=1.67$ GeV. One observes that $g_1$ and $g_T$ mostly differ in the positive-$x$ region. While $g_1$ has a smaller value than $g_T$ in the low-$x$ region, it has a much smaller slope at $x\approx0.1-0.3$. As a consequence, it becomes dominant in the region $0.2 \lesssim x \lesssim 0.5$. The two distributions are in agreement in the antiquark region within uncertainties and the large positive $x$ region. We note that the quasi-distributions do not have the canonical support, while the distribution after the matching is found to vanish outside of the physical region.

The light-cone PDFs $g_T$ and $g_1$ (and the corresponding quasi-PDFs) are connected via the Burkhardt-Cottingham sum rule~\cite{Burkhardt:1970ti},
\begin{equation}
\label{eq:SR}
\int_{-1}^{1} dx\,g_1(x) - \int_{-1}^{1} dx\,g_T(x) = 0\,,
\end{equation}
which serves as an important check for the lattice data. We find, without any additional input, that Eq.~(\ref{eq:SR}) gives 0.01(20) and therefore, the sum rule is satisfied. This check also suggests that effects due to operator mixing could be relatively small.

\vspace*{0.25cm}
\noindent \textit{Wandzura-Wilczek approximation:}
The lattice data of this work for $g_T$ and $g_1$ may be used to test the WW approximation~\cite{Wandzura:1977qf}. For the first time, we present a check of the full $x$-dependence of the WW approximation in lattice QCD. 
As already discussed above, in this approximation, the twist-3 $g_T(x)$ is fully determined by the twist-2 $g_1(x)$ (and denoted by $g_T^{\rm WW}$),
\begin{equation}
\label{eq:gT_WW}
g_T^{\rm WW}(x)=\int_x^1 \frac{dy}{y} g_1(y)\,.
\end{equation}
We evaluate $g_T^{\rm WW}$ using the lattice data for a wide range of $x$. The resulting $x$-dependence can be compared to the data for $g_T(x)$, as shown in Fig.~\ref{fig:gT_WW} for $P_3=1.67$ GeV. A similar study for all momenta shows compatibility of $g_T^{\rm WW}$ in all $x$ regions for all momenta (see Fig.~S3 of the supplementary document). The focus is on the quark region ($x>0$), which is less susceptible to systematic uncertainties, as compared to the antiquark region. 
We find that for a considerable $x$-range, our numerical results for $g_T(x)$ are consistent with $g_T^{\rm WW}(x)$.
However, given the uncertainties of the final distributions, a violation of the WW approximation is still possible at the level of up to 40\% for $x\lesssim0.4$.
Interestingly, a check of the WW approximation based on experimental data leads to a similar possible violation at the level of $15-40\%$, depending on $x$ \cite{Accardi:2009au}.
It is also notable to mention that while the slopes of $g_T$ and $g_1$ differ (see Fig.~\ref{fig:gT_g1}), the slopes of $g_T$ and $g_T^{\rm WW}$ are the same up to $x\approx0.4$. It should be noted that the distribution functions in the small-$x$ region ($x\lesssim0.1$) cannot be extracted reliably from the current lattice parameters due to enhanced higher-twist effects. The same holds for the large-$x$ region. For more details, see Ref.~\cite{Cichy:2018mum}. 
The difference of $g_T$ and $g_T^{\rm WW}$ for large $x$ could be due to systematic uncertainties, yet to be investigated. 
However, it may also indicate larger violations of the WW approximation in this region.

As an additional consistency check, we calculate the r.h.s. of Eq.~(\ref{eq:gT_WW}) using $g_1$ from global fits by the NNPDF \cite{Nocera:2014gqa} and JAM \cite{Ethier:2017zbq} collaborations.
We find good agreement with lattice-extracted $g_T^{\rm WW}$ up to $x\approx 0.3$.
Above this $x$ value, the discrepancy again indicates possible systematic effects.

\begin{figure}[h!]
\includegraphics[scale=0.55]{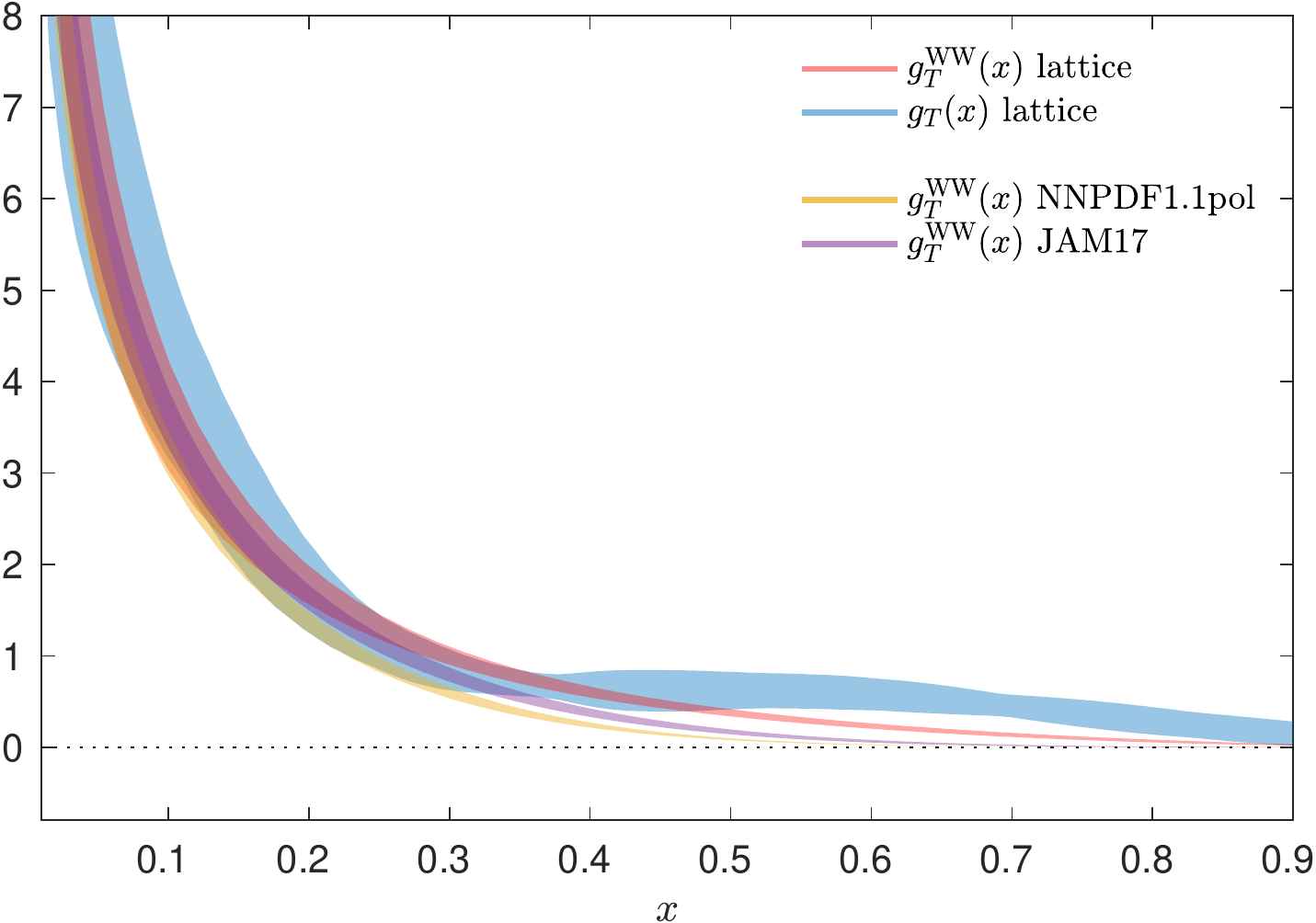}
	\caption{Comparison of our $g_T(x)$ (blue band) with its WW approximations: lattice-extracted $g_T^{\rm WW}$ (red band) and calculated from global fits (NNPDF1.1pol~\cite{Nocera:2014gqa} orange band, JAM17~\cite{Ethier:2017zbq} purple band). The proton momentum is $P_3=1.67$ GeV.}
 	\label{fig:gT_WW}
 \end{figure}

\vspace*{0.25cm}
\noindent \textit{Summary and prospects:}

We presented a pioneering \textit{ab initio} calculation of the proton twist-3 distribution $g_T(x)$, using numerical simulations of lattice QCD, within the quasi-distribution method. The work comprised multiple non-trivial steps, i.e.\ calculation of matrix elements of fast-moving protons and non-local operators in position space, elimination of divergences, reconstruction of the $x$-dependence, as well as matching to the light-cone distribution. For the quasi-distribution reconstruction, we used the Backus-Gilbert method, which improves the results by providing a unique solution to the inverse problem. 
Another novel result of this work is the calculation of a matching kernel for the case of $g_T$. Details on the extraction of the matching formula can be found in Ref.~\cite{Bhattacharya:2020xlt}.

The light-cone $g_T$ was obtained for three values of the momentum boost, $P_3=0.83,\,1.25,\,1.67$ GeV, and is presented in Fig.~\ref{fig:gTfinal}. We found that $g_T$ decays much faster than the leading-twist $g_1$, as shown in Fig.~\ref{fig:gT_g1}. 
A critical aspect of this work is the implementation of the Wandzura-Wilczek approximation, using both lattice data and data from global fits. 
We find $g_T$ consistent with its WW approximation for $x\lesssim0.4$, but within uncertainties, one can not exclude its violation at the level of up to 40\%, which is consistent with earlier studies based on experimental data.
A possibly larger violation is conceivable at larger $x$ according to our results.
Nevertheless, careful investigation of systematic uncertainties is needed for more precise quantitative statements, particularly at high-$x$.
The role of systematics in this region is confirmed by our consistency check comparing lattice-extracted $g_T^{\rm WW}$ with the ones from global fits, where agreement is observed for $x\lesssim0.3$.

We are considering several directions to extend this calculation.
Detailed investigations are required to quantify systematic uncertainties, such as excited states contamination, reconstruction of $x$-dependence, finite volume, and discretization effects. The latter two require a minimum of two and three ensembles, respectively. Also, simulations with quark masses fixed to their physical values (physical point) are now feasible within the available computational resources. We will move in that direction once systematic uncertainties for twist-3 distributions are understood. Operator mixing must be studied as well, which requires a new level of analytical and numerical work.

Finally, the possible breaking of the WW approximation at large $x$ signals a sizeable contribution from the $d_n$ terms. The connection between lattice estimates and results from experiments and phenomenology is more immediate for these quantities because of recent
measurements (see, e.g., Ref.~\cite{Flay:2016wie}) of $d_2$. The latter has a semiclassical interpretation of the average transverse force acting on the struck quark in a transversely polarized proton in DIS, right after it has been hit by the virtual photon~\cite{Burkardt:2008ps}. Our work generally shows that the calculation of the poorly known twist-3 PDFs of the proton within lattice QCD is within reach. We anticipate that the present study will stimulate further investigations in that area.

\begin{acknowledgements}
M.C. would like to thank Matthias Burkardt for useful discussions. The work of S.B.~and A.M.~has been supported by the National Science Foundation under grant number PHY-1812359.  A.M.~has also been supported by the U.S. Department of Energy, Office of Science, Office of Nuclear Physics, within the framework of the TMD Topical Collaboration. K.C.\ and A.S.\ are supported by the National Science Centre (Poland) grant SONATA BIS no.\ 2016/22/E/ST2/00013. F.S.\ was funded by DFG project number 392578569. M.C. acknowledges financial support by the U.S. Department of Energy, Office of Nuclear Physics, Early Career Award under Grant No.\ DE-SC0020405. Computations for this work were carried out in part on facilities of the USQCD Collaboration, which are funded by the Office of Science of the U.S. Department of Energy. 
This research was supported in part by PLGrid Infrastructure (Prometheus supercomputer at AGH Cyfronet in Cracow).
Computations were also partially performed at the Poznan Supercomputing and Networking Center (Eagle supercomputer), the Interdisciplinary Centre for Mathematical and Computational Modelling of the Warsaw University (Okeanos supercomputer), and at the Academic Computer Centre in Gda\'nsk (Tryton supercomputer). The gauge configurations have been generated by the Extended Twisted Mass Collaboration on the KNL (A2) Partition of Marconi at CINECA, through the Prace project Pra13\_3304 ``SIMPHYS".
\end{acknowledgements}

\newpage
\begin{widetext}

\setcounter{equation}{0}
\setcounter{figure}{0}
\setcounter{table}{0}

\parskip=6pt

{\bf \centerline{SUPPLEMENTARY MATERIAL}}

\bibliographystyle{apsrev}

\makeatletter
\renewcommand{\theequation}{S\arabic{equation}}
\renewcommand{\thefigure}{S\arabic{figure}}
\renewcommand{\thetable}{S\Roman{table}}

\section{Lattice calculation}
The main ingredient of this lattice calculation is matrix elements of the nonlocal operator
\begin{equation}
\label{eq:ME}
{\cal M}_{g_T}(P,z)\,=\,\langle P\vert \, \overline{\psi}(0,z)\,\gamma^j\,\gamma^5\, W(z)\,\psi(0,0)\,\vert P\rangle\,.
\end{equation}
where the fields are separated by a spatial distance $z$, which also indicates the direction of the Wilson line $W(z)$. The proton state is boosted with $\vec{P}=(0,0,P_3)$. We are interested in the isovector combination, which requires calculation of only the connected diagram shown in Fig.~\ref{fig:diagram}.
\begin{figure}[h]
\centering
\centerline{\includegraphics[scale=0.5]{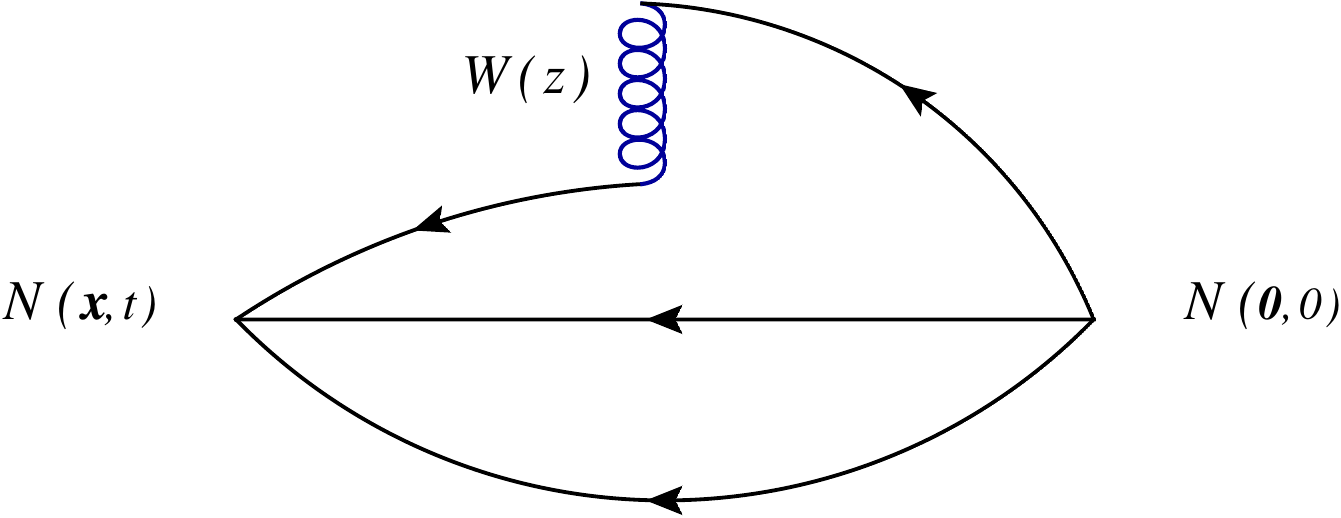}}
\vspace*{-0.35cm}
\caption{Diagram entering the calculation of the isovector combination for the three-point function. The initial and final states for the proton are indicated by $N(\mathbf{0},0)$ and $N(\mathbf{x},0)$, respectively.}
\label{fig:diagram}
\end{figure}

To obtain the proper contribution to $g_T$ from Eq.~(\ref{eq:ME}), we form an optimized ratio of three- and two-point correlation functions, that is
\be
C^{\rm 2pt}(\mathbf{P},t,0) =\Gamma_{\alpha\beta}\sum_\mathbf{x}e^{-i\mathbf{P}\cdot \mathbf{x}}\langle 0\vert N_\alpha(\mathbf{x},t) N_\beta(\mathbf{0},0)\vert 0 \rangle\,, 
\ee
\be
C^{\rm 3pt}(\mathbf{P};t,\tau,0) =  \Gamma'_{\alpha\beta}\,\sum_{\mathbf{x},\mathbf{y}}\,e^{-i\mathbf{P}\cdot \mathbf{x}} 
 \langle 0\vert N_{\alpha}(\mathbf{x},t) \mathcal{O}_{\gamma^j\,\gamma^5}(\mathbf{y},\tau;z)N_{\beta}(\mathbf{0},0)\vert 0\rangle\,.
\ee
$N_{\alpha}(x){=}\epsilon ^{abc}u^a _\alpha(x)\left( d^{b^{T}}(x)\mathcal{C}\gamma_5u^c(x)\right)$ is the interpolating field for the proton, $\tau$ is the current insertion time, and $\Gamma_{\alpha\beta}$ and $\Gamma'_{\alpha\beta}$ are 
the parity projector. $\Gamma_{\alpha\beta}$ is the parity plus projector
$\Gamma{=}\frac{1{+}\gamma_4}{2}$, and $\Gamma_{\alpha'\beta'}=\frac{1}{4}(1+\gamma^0) i \gamma^5 \gamma^j$. The ground state contribution is obtained from a constant fit in the plateau region of the ratio three- over two-point functions.

To increase the overlap with the proton ground state, we apply Gaussian smearing~\cite{Gusken:1989qx,Alexandrou:1992ti} and APE smearing~\cite{Albanese:1987ds} on the initial and final states (50 iterations each). The parameters are $\alpha_G{=}4$ for the Gaussian smearing and $\alpha_{APE}{=}0.5$ for APE smearing. We also employ the momentum smearing technique~\cite{Bali:2016lva}, which was proven to be crucial to suppress statistical uncertainties for matrix elements with boosted hadrons, and in particular, for nonlocal operators~\cite{Alexandrou:2016jqi}. Due to this method, we can reach a momentum boost of 1.67 GeV at a reasonable computational cost.

In Fig.~\ref{fig:bareME} we show the bare matrix elements $F_{g_T}$ at each value of $z/a$, and for the three values of $P_3$ used in this work. The top (bottom) plot shows the real (imaginary) part. Note that we focus on $z\ge0$, as the real (imaginary) part is an even (odd) function of $z$. 
 For both parts, we find convergence between the two largest momenta, within statistical uncertainties. However, there is no guarantee that the convergence will hold for $g_T(x)$, as the matching formula depends on $P_3$.

   \begin{figure}[h!]
 	\centering 
 	\hspace*{-2mm}
    \includegraphics[scale=0.58]{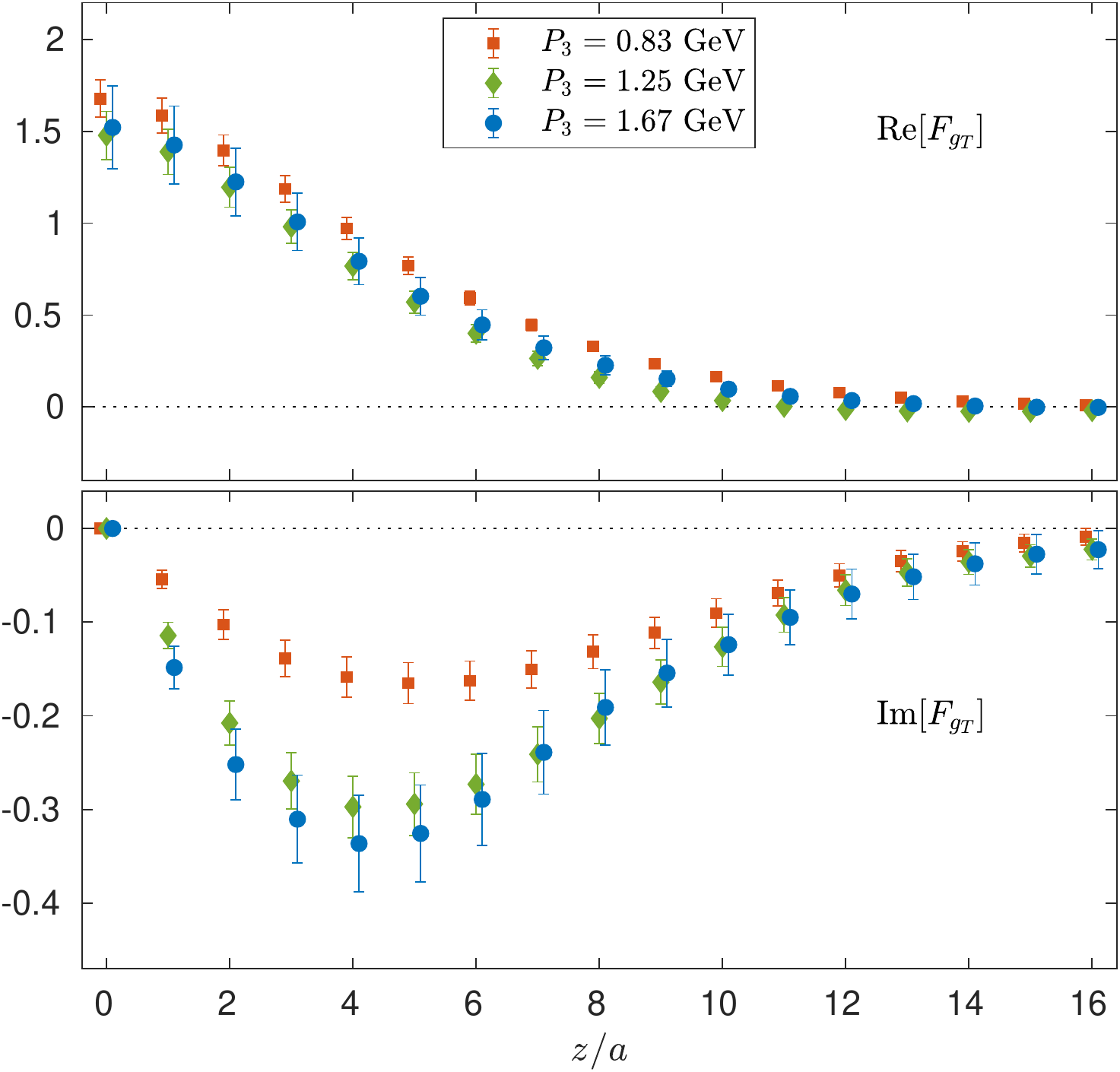}
 	\caption{Real (top) and imaginary (bottom) parts of the isovector bare matrix element ${F}_{g_T}$ for $P_3=0.83,\,1.25,\,1.67$ GeV, shown with red squares, green diamonds, blue circles, respectively.}
 	\label{fig:bareME}
 \end{figure}

\vspace*{0.25cm}
The renormalization procedure is performed non-perturbatively in the RI$'$-type scheme~\cite{Martinelli:1994ty}, using the momentum source method~\cite{Gockeler:1998ye,Alexandrou:2015sea} that offers high statistical accuracy. The appropriate conditions for the renormalization functions of the nonlocal operator, $Z_{g_T}$, and the quark field, $Z_q$, are
\be
\label{renorm}
Z_q^{-1}Z_{g_T}(z)\,\frac{1}{12} {\rm Tr} \left[{\cal V}_{g_T}(p,z) \left({\cal V}_{g_T}^{\rm Born}(p,z)\right)^{-1}\right] \Bigr|_{p^2{=}\bar\mu_0^2} {=} 1\, , \qquad
Z_q \, \frac{1}{12} {\rm Tr} \left[(S(p))^{-1}\, S^{\rm Born}(p)\right] \Bigr|_{p^2=\bar\mu_0^2}\,.
\ee
Note that the first equation is applied at each value of $z$ separately. ${\cal V}(p,z)$ ($S(p)$) is the amputated vertex function of the operator (fermion propagator) and ${\cal V}^{{\rm Born}}$ ($S^{{\rm Born}}(p)$) is its tree-level value. 

We calculate $Z_{g_T}$ using the five $N_f=4$ ensembles given in Tab.~\ref{Table:Z_ensembles}. These gauge configurations are dedicated to the calculation of the renormalization functions. Therefore, they correspond to the same $\beta$ value as for the $N_f=2+1+1$ ensemble used for the extraction of $g_T$. The need for degenerate quark is for a proper chiral extrapolation.
\begin{table}[h]
\begin{center}
\renewcommand{\arraystretch}{1.5}
\renewcommand{\tabcolsep}{5.5pt}
\begin{tabular}{ccc}
\hline\hline 
$\beta=1.726$, & $c_{\rm SW} = 1.74$, & $a=0.093$~fm \\
\hline\hline\\[-3ex]
{$24^3\times 48$}  & {$\,\,a\mu = 0.0060$}  & $\,\,m_\pi = 357.84$~MeV     \\
\hline
{$24^3\times 48$}  & $\,\,a\mu = 0.0080$     & $\,\,m_\pi = 408.11$~MeV     \\
\hline
{$24^3\times 48$}  & $\,\,a\mu = 0.0100$    & $\,\,m_\pi = 453.48$~MeV    \\
\hline
{$24^3\times 48$}  & $\,\,a\mu = 0.0115$    & $\,\,m_\pi = 488.41$~MeV    \\
\hline
{$24^3\times 48$}  & $\,\,a\mu = 0.0130$    & $\,\,m_\pi = 518.02$~MeV    \\
\hline\hline
\end{tabular}
\vspace*{-0.25cm}
\begin{center}
\caption{\small{Parameters of the $N_f=4$ ensembles used for the calculation and chiral extrapolation of the renormalization functions}.}
\label{Table:Z_ensembles}
\end{center}
\end{center}
\vspace*{-0.2cm}
\end{table} 

The RI$'$ renormalization scale $\bar\mu_0$ on which the renormalization functions are defined in Eq.~(\ref{renorm}) are carefully chosen to reduce discretization effects, as explained in Ref.~\cite{Alexandrou:2015sea}. Therefore, we choose the momentum of the vertex function to have the same spatial components, that is $p=(p_0,p_1,p_1,p_1)$, so that the ratio $\frac{p^4}{(p^2)^2}$ is less than 0.35, as suggested in Ref.~\cite{Constantinou:2010gr}. In this work, we use 17 different values of $\bar\mu_0$, between the range $(a\,\bar\mu_0)^2 \in [0.7,2.6]$. We apply a chiral extrapolation for each value of the  RI$'$ scale, using the fit
\begin{equation}
\label{eq:Zchiral_fit}
Z^{\rm RI}_{g_T}(z,\bar\mu_0,m_\pi) = {Z}^{\rm RI}_{g_T,0}(z,\mu_0) + m_\pi^2 \,{Z}^{\rm RI}_{gT,1}(z,\mu_0) \,,
\end{equation}
to extract the appropriate mass-independent estimate ${Z}^{\rm RI}_{g_T,0}(z,\mu_0)$. The chirally extrapolated values are converted to the $\overline{\rm MS}$ scheme and evolved to $\mu{=}2$ GeV using the results of Ref.~\cite{Constantinou:2017sej}. Finally, we obtain ${Z}^{\overline{\rm MS}}_{g_T,0}(z,2 {\rm GeV})$ by extrapolating $(a\,p)^2 {\to} 0$ using a linear fit and data in the region $(a\,\bar\mu_0)^2 \in [1,2.6]$.

\section{WW Approximation}
 
     \begin{figure}[h!]
 	\centering 
 	\hspace*{-2mm}
    \includegraphics[scale=0.65]{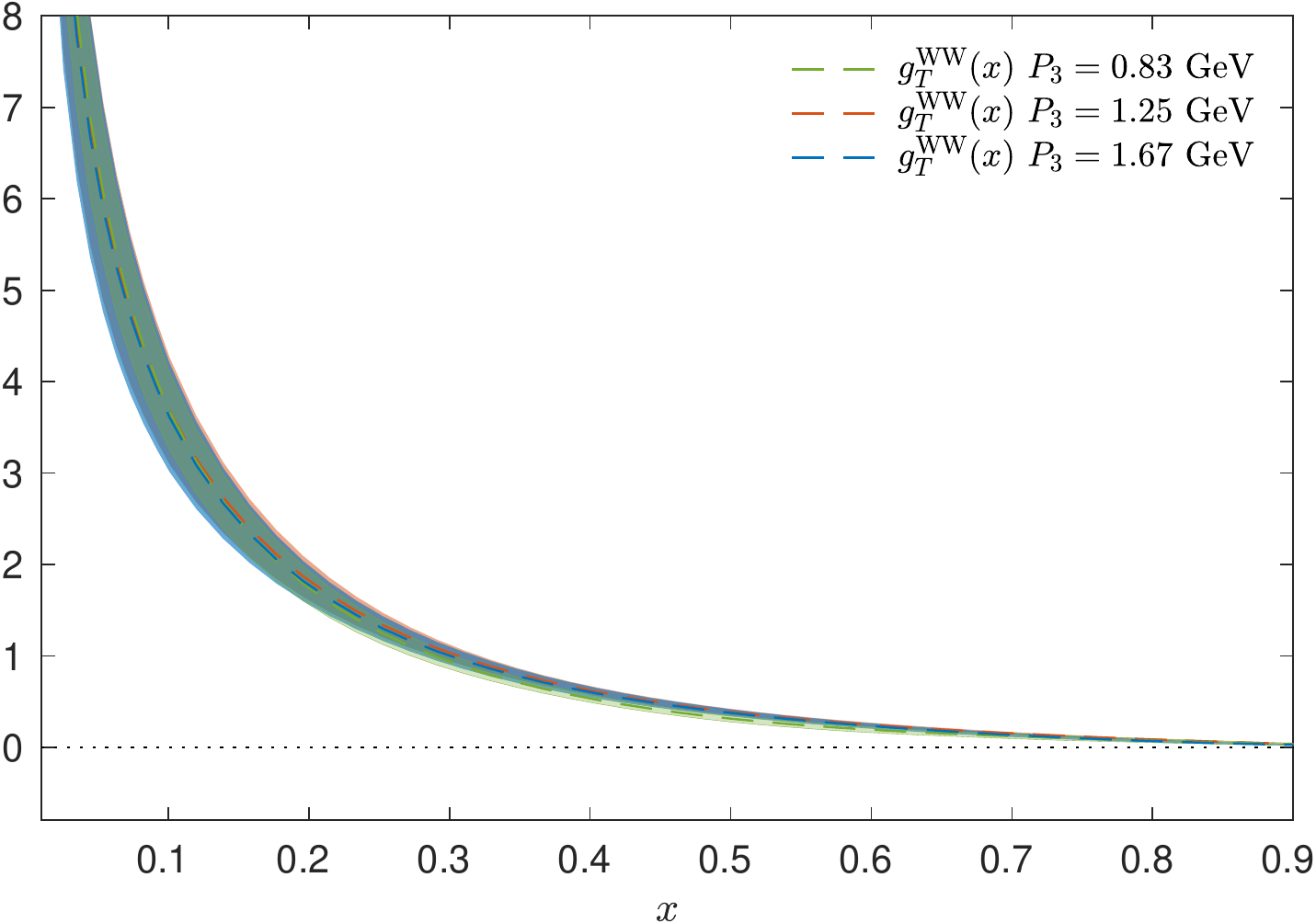}
 	\caption{$g_T^{\rm WW}$ obtained for $P_3=0.83,\,1.25,\,1.67$ GeV, with green, red, blue band, respectively.}
 	\label{fig:gT_WW}
 \end{figure}
 
 A useful check for the WW approximation is whether the conclusions described in the main text change with the value of $P_3$ used for the comparison. In Fig.~\ref{fig:gT_WW} we show the lattice results for $g_T^{\rm WW}$ obtained from the three momenta. We find that the three curves are fully compatible within the statistical uncertainties (width of bands), which indicates that convergence has been achieved for this lattice setup.

\end{widetext}

\bibliography{references}

\end{document}